\documentclass[%
 reprint,
superscriptaddress,
showpacs,
showkeys,
 amsmath,amssymb,
prl,
]{revtex4-2}

\usepackage{graphicx}
\usepackage{dcolumn}
\usepackage{bm}
\usepackage{esvect}
\usepackage{accents}

\begin{document}

\preprint{APS/123-QED}

\title{MICROSCOPE mission: final results of the test of the Equivalence Principle}

\author{Pierre Touboul}
\thanks{Deceased in February 2021}
\affiliation{ONERA, Universit\'e Paris Saclay, F-91123 Palaiseau, France}

 \author{Gilles M\'etris}
 \email{Gilles.Metris@oca.eu}
  \affiliation{Universit\'e C\^ote d'Azur, Observatoire de la C\^ote d'Azur, CNRS, IRD, G\'eoazur, 250 avenue Albert Einstein, F-06560 Valbonne, France}

\author{Manuel Rodrigues}
\email{manuel.rodrigues@onera.fr}
\affiliation{DPHY, ONERA, Universit\'e Paris Saclay, F-92322 Ch\^atillon, France}

\author{Joel Berg\'e}
\affiliation{DPHY, ONERA, Universit\'e Paris Saclay, F-92322 Ch\^atillon, France}

\author{Alain Robert}
\affiliation{CNES Toulouse, 18 avenue Edouard Belin - 31401 Toulouse Cedex 9, France}

\author{Quentin Baghi}
\altaffiliation{Current address: CEA, Centre de Saclay, IRFU/DPhP, 91191 Gif-sur-Yvette, France}
\affiliation{Universit\'e C\^ote d'Azur, Observatoire de la C\^ote d'Azur, CNRS, IRD, G\'eoazur, 250 avenue Albert Einstein, F-06560 Valbonne, France}
\affiliation{DPHY, ONERA, Universit\'e Paris Saclay, F-92322 Ch\^atillon, France}

\author{Yves Andr\'e}
\affiliation{CNES Toulouse, 18 avenue Edouard Belin - 31401 Toulouse Cedex 9, France}

\author{Judica\"el Bedouet}
\affiliation{ONERA, Universit\'e de Toulouse, F-31055 Toulouse, France}

\author{Damien Boulanger}
\affiliation{DPHY, ONERA, Universit\'e Paris Saclay, F-92322 Ch\^atillon, France}

\author{Stefanie Bremer}
\altaffiliation{Current address: DLR, Institute for Satellite Geodesy and Inertial Sensing, Am Fallturm 9, D-28359 Bremen, Germany}
\affiliation{ZARM, Center of Applied Space Technology and Microgravity, University of Bremen, Am Fallturm, D-28359 Bremen, Germany}

\author{Patrice Carle}
\affiliation{ONERA, Universit\'e Paris Saclay, F-91123 Palaiseau, France}

\author{Ratana Chhun}
\affiliation{DPHY, ONERA, Universit\'e Paris Saclay, F-92322 Ch\^atillon, France}

\author{Bruno Christophe}
\affiliation{DPHY, ONERA, Universit\'e Paris Saclay, F-92322 Ch\^atillon, France}

\author{Valerio Cipolla}
\affiliation{CNES Toulouse, 18 avenue Edouard Belin - 31401 Toulouse Cedex 9, France}

\author{Thibault Damour}
\affiliation{IHES, Institut des Hautes Etudes Scientifiques, 35 Route de Chartres, 91440 Bures-sur-Yvette, France}

\author{Pascale Danto}
\affiliation{CNES Toulouse, 18 avenue Edouard Belin - 31401 Toulouse Cedex 9, France}

\author{Louis Demange}
\affiliation{Universit\'e C\^ote d'Azur, Observatoire de la C\^ote d'Azur, CNRS, IRD, G\'eoazur, 250 avenue Albert Einstein, F-06560 Valbonne, France}

\author{Hansjoerg Dittus}
\affiliation{DLR, K\"oln headquarters, Linder H\"ohe, 51147 K\"oln, Germany}

\author{Oc\'eane Dhuicque}
\affiliation{DPHY, ONERA, Universit\'e Paris Saclay, F-92322 Ch\^atillon, France}

\author{Pierre Fayet}
\affiliation{Laboratoire de physique de l'Ecole normale sup\'erieure, ENS, Universit\'e PSL, CNRS, Sorbonne Universit\'e, Universit\'e de Paris, \mbox{F-75005 Paris, France,} and CPhT, Ecole polytechnique, IPP, Palaiseau, France}

\author{Bernard Foulon}
\affiliation{DPHY, ONERA, Universit\'e Paris Saclay, F-92322 Ch\^atillon, France}

\author{Pierre-Yves Guidotti}
\thanks{Current address: AIRBUS Defence and Space, F-31402 Toulouse, France}
\affiliation{CNES Toulouse, 18 avenue Edouard Belin - 31401 Toulouse Cedex 9, France}

\author{Daniel Hagedorn}
\affiliation{PTB, Physikalisch-Technische Bundesanstalt, Bundesallee 100, 38116 Braunschweig, Germany}

\author{Emilie Hardy}
\affiliation{DPHY, ONERA, Universit\'e Paris Saclay, F-92322 Ch\^atillon, France}

\author{Phuong-Anh Huynh}
\affiliation{DPHY, ONERA, Universit\'e Paris Saclay, F-92322 Ch\^atillon, France}

\author{Patrick Kayser}
\affiliation{DPHY, ONERA, Universit\'e Paris Saclay, F-92322 Ch\^atillon, France}

\author{St\'ephanie Lala}
\affiliation{ONERA, Universit\'e Paris Saclay, F-91123 Palaiseau, France}

\author{Claus L\"ammerzahl}
\affiliation{ZARM, Center of Applied Space Technology and Microgravity, University of Bremen, Am Fallturm, D-28359 Bremen, Germany}

\author{Vincent Lebat}
\affiliation{DPHY, ONERA, Universit\'e Paris Saclay, F-92322 Ch\^atillon, France}

\author{Fran\c{c}oise Liorzou}
\affiliation{DPHY, ONERA, Universit\'e Paris Saclay, F-92322 Ch\^atillon, France}

\author{Meike List}
\thanks{Current address: DLR, Institute for Satellite Geodesy and Inertial Sensing, Am Fallturm 9, D-28359 Bremen, Germany}
\affiliation{ZARM, Center of Applied Space Technology and Microgravity, University of Bremen, Am Fallturm, D-28359 Bremen, Germany}

\author{Frank L\"offler}
\affiliation{PTB, Physikalisch-Technische Bundesanstalt, Bundesallee 100, 38116 Braunschweig, Germany}

\author{Isabelle Panet}
\affiliation{IPGP, 35 rue H\'el\`ene Brion, 75013 Paris, France}

\author{Martin Pernot-Borr\`as}
\affiliation{DPHY, ONERA, Universit\'e Paris Saclay, F-92322 Ch\^atillon, France}

\author{Laurent Perraud}
\affiliation{CNES Toulouse, 18 avenue Edouard Belin - 31401 Toulouse Cedex 9, France}

\author{Sandrine Pires}
\affiliation{Universit\'e Paris Saclay et Universit\'e de Paris, CEA, CNRS, AIM, F-91190 Gif-sur-Yvette, France}

\author{Benjamin Pouilloux}
\altaffiliation{Current address: KINEIS, F-31520 Ramonville Saint-Agne, France}
\affiliation{CNES Toulouse, 18 avenue Edouard Belin - 31401 Toulouse Cedex 9, France}

\author{Pascal Prieur}
\affiliation{CNES Toulouse, 18 avenue Edouard Belin - 31401 Toulouse Cedex 9, France}

\author{Alexandre Rebray}
\affiliation{DPHY, ONERA, Universit\'e Paris Saclay, F-92322 Ch\^atillon, France}

\author{Serge Reynaud}
\affiliation{Laboratoire Kastler Brossel, Sorbonne Universit\'e, CNRS, ENS-PSL Universit\'e, Coll\`ege de France, 75252 Paris, France}

\author{Benny Rievers}
\affiliation{ZARM, Center of Applied Space Technology and Microgravity, University of Bremen, Am Fallturm, D-28359 Bremen, Germany}

\author{Hanns Selig}
\thanks{Current address: GERADTS GMBH, Kleiner Ort 8, D-28357 Bremen, Germany}
\affiliation{ZARM, Center of Applied Space Technology and Microgravity, University of Bremen, Am Fallturm, D-28359 Bremen, Germany}

\author{Laura Serron}
\affiliation{Universit\'e C\^ote d'Azur, Observatoire de la C\^ote d'Azur, CNRS, IRD, G\'eoazur, 250 avenue Albert Einstein, F-06560 Valbonne, France}

\author{Timothy Sumner}
\affiliation{Blackett Laboratory, Imperial College London, Prince Consort Road, London. SW7 2AZ, United Kingdom}

\author{Nicolas Tanguy}
\affiliation{DPHY, ONERA, Universit\'e Paris Saclay, F-92322 Ch\^atillon, France}

\author{Patrizia Torresi}
\affiliation{CNES Toulouse, 18 avenue Edouard Belin - 31401 Toulouse Cedex 9, France}

\author{Pieter Visser}
\affiliation{Faculty of Aerospace Engineering, Delft University of Technology, Kluyverweg 1, 2629 HS Delft, Netherlands}

\date{\today}

\begin{abstract}
The MICROSCOPE mission was designed to test the Weak Equivalence Principle (WEP), stating the equality between the inertial and the gravitational masses, with a precision of $10^{-15}$ in terms of the E\"otv\"os ratio $\eta$. Its experimental test consisted of comparing the accelerations undergone by two collocated test masses of different compositions as they orbited the Earth, by  measuring the electrostatic forces required to keep them in equilibrium. This was done with ultra-sensitive differential electrostatic accelerometers onboard a drag-free satellite. The mission lasted two and a half years, cumulating five-months-worth of science free-fall data, two thirds with a pair of test masses of different compositions -- Titanium and Platinum alloys -- and the last third with a reference pair of test masses of the same composition -- Platinum. We summarize the data analysis, with an emphasis on the characterization of the systematic uncertainties due to thermal instabilities and on the correction of short-lived events which could mimic a WEP violation signal. We found no violation of the WEP, with the E\"otv\"os parameter of the Titanium and Platinum pair constrained to $\eta({\rm Ti, Pt})~=~ [-1.5 \pm 2.3~{\rm (stat)} \pm 1.5~{\rm (syst)}]~\times 10^{-15}$ at $1\sigma$ in statistical errors.
\end{abstract}

\maketitle


General Relativity (GR) offers a remarkable description of gravitational interactions, successfully tested in the anomalous precession of the perihelion of Mercury, the bending of light in a gravitational field, the gravitational redshift, the Shapiro time delay and the change in the periods of binary pulsars from the emission of gravitational waves \cite{will14, will18, tino20, bertotti03, hoekstra08, fienga15, delva18, herrmann18, abuter20, kramer21}. Gravitational waves from the coalescence of neutron stars and very massive black holes have been observed recently, providing evidence for the existence of black holes and ruling out many beyond-GR models \cite{abbott16, abbott17, brax16, lombriser16, lombriser17, baker17, creminelli17, ezquiaga17, sakstein17}.

A building block of general relativity is the Equivalence Principle (EP), according to which all bodies fall in the same way in a gravitational field when no other forces are acting on them, independently of their masses and internal constitutions. First observed by Galileo and Newton and tested by E\"otv\"os et al. at the $5\times 10^{-9}$ level \cite{eotvos22}, the universality of free fall was elevated to a principle by Einstein, the weak equivalence principle (WEP), taken as a cornerstone of general relativity \cite{einstein07}.

Still the above tests of GR are classical, i.e. not involving quantum physics. 
But one does not know how to cast GR into a consistent quantum theory, even if several approaches, including most notably string theories \cite{becker06}, have been developed to tackle this problem. At the conceptual level, also, physicists have been dreaming of a unified theory including strong, electromagnetic and weak interactions as well as gravity.

As GR cannot be considered as a complete theory, and in view of other questions such as the nature of dark energy \cite{riess98, perlmutter99} and dark matter \cite{zwicky33, rubin70}, it is important to test as precisely as possible the EP. Both the need to complete GR if it is to be turned into a satisfactory quantum theory, and the desire for a unified description of interactions, lead to consider the possibility of new long-ranged interactions and forces, that could lead to very small apparent violations of the EP. 

Thereby, a test of the EP appears as a test of one of the basic principles of GR, and also serves as a search for new interactions. While gravity is supposed to be mediated by the (still hypothetical) spin-2 graviton, extremely weak new forces could be mediated by very light or massless spin-0 or spin-1 bosons, that may be thought of as part of a completion of GR. Such EP-violating spin-0 bosons, like the dilaton or other dilaton-like particles, tend to appear within string theories \cite{damour94, damour02, khoury04a, khoury04b, clifton12, joyce15, vainshtein72, babichev09, hinterbichler10, brax13, burrage18}. A spin-1 boson U associated with an extension of the standard model gauge group is expected to couple to a combination of baryonic, leptonic (or B-L within grand-unification) and electromagnetic currents (with possibly axial couplings, of no effects here) \cite{fayet89, fayet90}. 

The WEP has been intensively tested throughout the past four centuries \cite{fischbach99, will14, eotvos22, roll64, dicke70, braginsky71}, and verified to a precision of $2~\times~ 10^{-13}$ in the first decade of the 21st century \cite{schlamminger08, wagner12, williams12}. This precision has been increased by one order of magnitude ($2~\times~ 10^{-14}$) with the first results of the MICROSCOPE mission in 2017 \cite{touboul17, touboul19}, taking advantage of space quietness \cite{chapman01} and of new instrument capabilities \cite{everitt03}. At about the same time, over 48 years of lunar laser ranging (LLR) data allowed for a $7.1\times10^{-14}$ precision \cite{viswanathan18}. We report here the final MICROSCOPE mission results, setting the tightest bound on the validity of the WEP achieved to date, also providing improved constraints on additional new forces \cite{berge18, fayet18, fayet19}.

\bigskip
The MICROSCOPE space mission was designed following developments of the Satellite Test of the Equivalence Principle (STEP \cite{mester01, sumner07}), built from Chapman's seminal proposal to test the WEP in the Earth orbit as early as the 1970s \cite{chapman01}.
Albeit less ambitious than STEP, it was designed to test the WEP in space in terms of the E\"otv\"os ratio
\begin{eqnarray} \label{eq_eta}
\eta_{\rm A,B}&=&2\frac{a_{\rm A}-a_{\rm B}}{a_{\rm A}+a_{\rm B}}  \\
&\simeq & \left(\frac{m_g}{m_i}\right)_{\rm A} - \left(\frac{m_g}{m_i}\right)_{\rm B} = \delta{\rm(A,B)}, \label{eq_delta}
\end{eqnarray}
where $a_{\rm A}$ and $a_{\rm B}$ are the accelerations of two free-falling test bodies A and B, and $m_g$ and $m_i$ their gravitational and inertial masses, respectively. Eq. (\ref{eq_delta}) defines the approximated E\"otv\"os ratio to be estimated by MICROSCOPE \cite{touboul09, touboul12}. The bodies are two concentric hollow cylindrical test masses controlled with electrostatic forces in a differential accelerometer. Any difference in the forces required to keep the two test masses in relative equilibrium would provide evidence for an apparent violation of the WEP, originating from an intrinsic violation, or as an effect of extremely small new forces \cite{rodriguescqg1}. 
MICROSCOPE includes two such differential accelerometers called sensor units: in the first one (SUREF), the two test masses have the same composition (Pt:Rh alloy); in the second one (SUEP), they have different compositions (PtRh(90/10) and TiAlV(90/6/4) alloys). The former serves as a reference instrument, while the latter is used to test the WEP. The test masses' characteristics are summarized in Table \ref{tab:masses}, and details about the instrument can be found in Refs. \cite{touboul17, touboul19, liorzoucqg2}.
The payload was integrated in a drag-free CNES microsatellite able to provide the experiment with a very quiet environment \cite{robertcqg3}. MICROSCOPE was launched from Kourou on April 25, 2016 and set into a sun-synchronous, dawn-dusk orbit to optimize its thermal stability. The mission ended on October 18, 2018. Ref. \cite{rodriguescqg4} presents the mission scenario.

\begin{table}[tb]
\caption{\label{tab:masses}%
Main test-mass physical properties measured in the laboratory before integration in the instrument.
}
\begin{ruledtabular}
\begin{tabular}{lcccc}
\textrm{Measured}&
\textrm{SUREF}&
\textrm{SUREF}&
\textrm{SUEP} & SUEP\\
parameters & Inner mass & Outer mass & Inner mass & Outer mass \\
at $20\,^o$C & Pt/Rh & Pt/Rh & Pt/Rh & Ti/Al/V \\
\hline

Mass in kg & 0.401533 & 1.359813 & 0.401706 & 0.300939\\ \hline
Density in  & 19.967 & 19.980 & 19.972 & 4.420\\
g\,cm$^{-3}$ &  &  &  & \\
\end{tabular}
\end{ruledtabular}
\end{table}

The experimental observable relevant to the test of the WEP is the difference between the electrostatic accelerations exerted on the inner (labeled 1) and the outer (labeled 2) test-masses of a given sensor unit ${\vec \Gamma_d}^{\rm meas}~ \equiv ~{\vec \Gamma_1}^{\rm meas} ~-~ {\vec \Gamma_2}^{\rm meas}$.
It is directly related to the E\"otv\"os ratio $\eta(2,1) ~\approx ~\delta(2,1)$ and to the various forces acting on the satellite (see Ref. \cite{rodriguescqg1} for a detailed derivation). In the instrument's reference frame,
\begin{multline} \label{eq_meas}
\overrightarrow\Gamma_d^{\rm meas} \simeq  \overrightarrow K_{0,d} \\ 
 +\left[A_c\right]\left\{ \left( \left[ T \right]- \left[{\rm In}\right]  \right)\overrightarrow\Delta - 2\left[\Omega \right] \dot{\overrightarrow\Delta}-\ddot{\overrightarrow\Delta} +\delta\left(2,1\right) \overrightarrow g_{\rm sat} \right\}\\ 
 + 2\left[ A_d\right] \overrightarrow\Gamma_c^{\rm app} + 2 \left[ \rm C_d\right] \dot{\overrightarrow \Omega}+ \overrightarrow{n}_d,
\end{multline}
where $\overrightarrow K_{0,d}$ is a differential bias, $\overrightarrow\Delta$ connects the center of the inner mass to that of the outer mass, $[T]$ is the gravity gradient tensor in the satellite frame, $[{\rm In}]=[\dot{\overrightarrow \Omega}] + [\overrightarrow \Omega] [\overrightarrow \Omega]$ is the gradient of inertia, with $[\overrightarrow \Omega]$ the satellite's angular velocity matrix, $\overrightarrow g_{\rm sat}$ is the Earth gravity acceleration at the center of the satellite, $\overrightarrow\Gamma_c^{\rm app}$ is the mean acceleration applied on both masses, $\left[ \rm C_d\right]$ is the differential-mode linear-to-angular acceleration coupling matrix, $\overrightarrow{n}_d$ is the noise and dots denote differentiation with respect to time. Finally, the common- and differential-mode sensitivity matrices $[A_c]$ and $[A_d]$ are defined from the instrument's scale factors and test-mass reference frame defects. Those parameters are more fully described in Table II of Ref. \cite{touboul17} and in Refs. \cite{touboul19, rodriguescqg1, hardycqg6}. 

The test of the WEP is performed along the longitudinal axis of the test masses, designed to be the most sensitive. Data analysis thus deals with the differential acceleration (\ref{eq_meas}) projected along the longitudinal, sensitive $x$ axis of the instrument.
Ref. \cite{rodriguescqg1} provides the measurement equation projected on this axis.

The satellite can be spun around the normal $y$-axis to the orbital plane and oppositely to the orbital motion in order to increase the frequency of the Earth
gravity modulation. In this case, in the satellite frame, the Earth gravity field rotates at the sum of the orbital and spin frequencies. A WEP violation would give a signal modulated at this frequency, denoted $f_{\rm EP}$. The frequencies used during the mission are listed in Table \ref{tab_freqs}. Ref. \cite{rodriguescqg4} introduces the concept of sessions: WEP test sessions last several days and are defined by their spin frequencies, while calibration sessions are short and allow us to estimate instrumental parameters.

\begin{table}[tb]
\caption{\label{tab_freqs}%
Frequencies of interest.
}
\begin{ruledtabular}
\begin{tabular}{lcc}
& Frequency [$\times 10^{-3}$ Hz] & Comment \\
\hline
$f_{\rm{orb}}$ & 0.16818 & Mean orbital frequency \\
$f_{\rm{spin}_2}$ & $\frac{9}{2}f_{\rm{orb}}=0.75681$ & Spin rate frequency 2 (V2 mode) \\
$f_{\rm{spin}_3}$ & $\frac{35}{2} f_{\rm{orb}}=2.94315$ & Spin rate frequency 3 (V3 mode) \\
$f_{\rm{EP}_2}$ & 0.92499 & EP frequency in V2 mode  \\
$f_{\rm{EP}_3}$ & 3.11133 & EP frequency in V3 mode \\
$f_{\rm cal}$ & $1.22848$ & Calibration frequency \\
\end{tabular}
\end{ruledtabular}
\end{table}

The analysis of first results in Refs. \cite{touboul17, touboul19} used one 120-orbit session on SUEP to obtain $\delta({\rm Ti,Pt})=[-1\pm{}9{\rm (stat)}\pm{}9{\rm (syst)}] \times{}10^{-15}$
at 1$\sigma$ statistical uncertainty.
No calibration was used, and systematic errors were dominated by thermal effects, for which an upper bound was used. Improvements were then expected in the pursuit of the mission and its data analysis.

We report here the final results of the MICROSCOPE mission, based on eighteen sessions for SUEP and nine sessions for SUREF, with all data calibrated and systematics now fully characterized \cite{chhuncqg5, hardycqg6}. Ref. \cite{bergecqg7} presents the methods used for data analysis, and Ref. \cite{metriscqg9} details their results on the actual data. The main aspects of the data and its analysis are summarized below.

\bigskip
A handful of sessions were discarded because of non-linearities at the beginning of the mission, before the control loop's electronics was upgraded. A few others were discarded because of rare anomalies. The results presented in this paper were then obtained from eighteen sessions on SUEP  and nine on SUREF, including two with both SUs switched on together \cite{metriscqg9}. 
Hereafter, we first describe the data analysis methods before discussing the actual in-flight estimation of instrumental parameters and presenting the E\"otv\"os parameter determination.

Each in-flight calibration session is dedicated to estimating one or two parameters and designed so that the signals sourced by those parameters have a favourable signal-to-noise ratio.
Using a Markov Chain Monte Carlo (MCMC) technique on nine sessions, we showed that it is possible to cumulate sessions and estimate all parameters simultaneously from Eq. (\ref{eq_meas}). This unpublished study, based on Ref. \cite{vitale14}, used the data of Ref. \cite{touboul17} and gave results consistent with those of the technique described below. Indeed, instead of using a CPU-expensive MCMC method, we use the fact that parameters are almost independent to simplify and better control the estimation process. This is done via the following iterative method based on the {\sc Adam} (Accelerometric Data Analysis for MICROSCOPE) code to estimate parameters in the frequency domain. The method is presented and applied to numerical simulations in Ref. \cite{bergecqg7}.

When projected on the $x$-axis, the measurement equation (\ref{eq_meas}) is of the form $\Gamma_{d,x} ~=~ f(p_k, t) ~+~ n_{d,x}$, where $p_k$ are parameters and the time dependence is related to measured or modeled signals $s_i(t)$. For each session, the data provides us with $\Gamma_{d,x}$ and all $s_i(t)$, allowing for the estimation of the parameters $p_k$.
Moreover, {\it a priori} values $p_{k,0}$ (either measured on ground or estimated during an earlier in-flight calibration session) are used to correct the measurement for the corresponding signals and to refine the estimation of some parameters $p_{ke}$.

In practice, instrumental defects are parameterized by the $\overrightarrow K_{0,d}$ and $\overrightarrow\Delta$ vectors, as well as the $\left[A_d\right]$, $\left[A_c\right]$ and $\left[ \rm C_d\right]$ matrices in Eq. (\ref{eq_meas}). Note that only some of their components impact the projected acceleration. Ref. \cite{rodriguescqg1} details how they affect the measurement and Ref. \cite{hardycqg6} shows how they were estimated in flight and evolve in time. 
The estimation of $\Delta_x$ and $\Delta_z$ takes advantage of their couplings with the Earth gravity gradient, whose strong line at $2f_{\rm EP}$ allows for a direct determination in science data from an accurate Earth gravity model \cite{metris98, mayer06}. Dedicated 5-orbit sessions were used to measure $\Delta_y$, where the satellite was oscillated about the $z$-axis at frequency $f_{\rm cal}$ to create a measurable signal driven by $\Delta_y$ at $f_{\rm cal}$. The elements of the first row of the $[A_d]$ matrix $a_{d1i}$ were measured by shaking the satellite at frequency $f_{\rm cal}$ along each axis ($x$ to measure $a_{d11}$, $y$ for $a_{d12}$ and $z$ for $a_{d13}$) in order to drive a measurable signal dependent on those parameters. The $a_{d11}$ sessions also allowed for a measurement of the differential quadratic factor $K_{2d,xx}$ at $2f_{\rm cal}$. Although we found a slight correlation of $\Delta_x$ and $\Delta_z$ with temperature, the other parameters remained roughly constant during the mission \cite{hardycqg6}. Table \ref{tab_params} lists their mean values.

\begin{table}[tb]
\caption{\label{tab_params}%
Mean estimated values of the off-centerings' components and of the first row of the $[A_d]$ matrix estimated in flight \cite{hardycqg6}. The off-centerings components $\Delta_x$ and $\Delta_z$ correspond to sessions within a limited temperature range.}
\begin{ruledtabular}
\begin{tabular}{lcc}
& SUEP & SUREF \\
\hline
$\Delta_x~[\mu$m] & $19.998\pm0.009$ & $-35.884\pm0.005$\\
$\Delta_y~[\mu$m] & $-8.19\pm0.09$ & $5.89\pm0.05$ \\
$\Delta_z~[\mu$m] & $-5.605\pm0.009$ & $5.712\pm0.005$\\
$a_{d11}$ [$\times 10^{-3}$] & $8.5\pm0.2$ & $-14.6\pm0.2$\\
$a_{d12}$ [$\times 10^{-5}~{\rm mrad}$] & $-25.6\pm0.5$ & $-3.5\pm1.5$\\
$a_{d13}$ [$\times 10^{-5}~{\rm mrad}$] & $13.6\pm0.9$ & $-9.1\pm0.3$\\
$K_{2d,xx}$ & $-1037\pm4800$ & $2409\pm1650$
\end{tabular}
\end{ruledtabular}
\end{table}

Once the above iterative process had converged (typically in two to three iterations), we estimated the E\"otv\"os parameter on calibrated data following the corrected measurement equation
\begin{equation}
  \label{eq_xaccEP}
  \Gamma_{d,x}^{\rm corr}= \tilde{b}_{d,x}'+\delta_x g_x+\delta_z  g_z+\Delta_{x}  S_{xx} +\Delta_{z}  S_{xz}+ n_{d, x},
\end{equation}
which is the core model fitted to the data, where $\tilde{b}_{d,x}'$ is the bias. In addition to the E\"otv\"os parameter  $\delta_x $ we also estimated the amplitude $\delta_z$ of a signal proportional to $g_z$ (varying also at $f_{\rm EP}$ but in quadrature with $g_x$) and the components $\Delta_{x}$ and $\Delta_{z}$ of the apparent offcentering.

Additionally, spurious events can be spotted in the data. ``Glitches" are short-lived events, most probably due to crackles of the satellite's multi-layer insulator \cite{bergecqg8}. 
Glitches occur quasi-periodically and can impart a signal at $f_{\rm EP}$. Although numerical models do not allow for the estimation of the level of this signal, we noticed that removing glitches from sessions with a strong signal at $f_{\rm EP}$ -- statistically inconsistent with other sessions -- decreases the signal, hinting at a significant effect from glitches on the E\"otv\"os parameter estimation. 
Therefore, to counteract their direct effect on MICROSCOPE's WEP measurement, we masked them as follows. We used a standard recursive $\sigma$-clipping technique -- $\sigma$ being the standard deviations of the data -- to search for outliers, defined as points that deviate by more than (i) $4.5~\sigma$ from the moving average of the data and (ii) more than $3~\sigma$ from the moving average of the high-frequency-filtered data. We then mask one (15) second(s) before (after) each outlier to make sure that the transient regime was always removed \cite{rodriguescqg4, metriscqg9}.
Masked glitches thus behaved as ``missing" data, so that data became unevenly sampled in time, thereby hampering {\sc Adam}'s fit in the frequency domain \cite{baghi15,berge15b}.

We tackled this difficulty with the M-ECM (Modified-Expectation-Conditional-Maximization \cite{baghi16}) technique; it maximizes the likelihood of available data through the estimation of missing data by their conditional expectation, based on the circulant approximation of the complete data covariance. We showed in Ref. \cite{baghi16} that it faithfully reconstructs the noise power spectral density and provides unbiased estimates of parameters. Finally, we added and correctly measured mock WEP violation signals in the data to make sure that this procedure does not affect a possible real WEP violation signal.

M-ECM also fills gaps, and we can then use {\sc Adam} to cross-check M-ECM's estimates of the E\"otv\"os parameter for each session. We also use it to combine all sessions and infer the overall constraint given below.

In addition to glitches, rare jumps in the differential acceleration can be spotted, mostly on SUREF \cite{metriscqg9}. These jumps are not simple discontinuities, but appear as unsteady transitions between two stable states. Although hidden in the noise, they perturb the data analysis and must be discarded. Since this amounts to creating gaps of several hundred seconds, we decided to extract ``segments" between jumps (or between jumps and any extremity of the session), when such jumps existed (otherwise, we call ``segment" the entire session). Segments are as long as possible and consist of an even number of orbital periods to ensure that potential contamination by signals at frequencies $mf_{\rm orb} + nf_{\rm spin}$ ($m,n \in \mathbb{N}$) are cancelled \cite{metriscqg9}.

\begin{figure}
\includegraphics[width=0.45\textwidth]{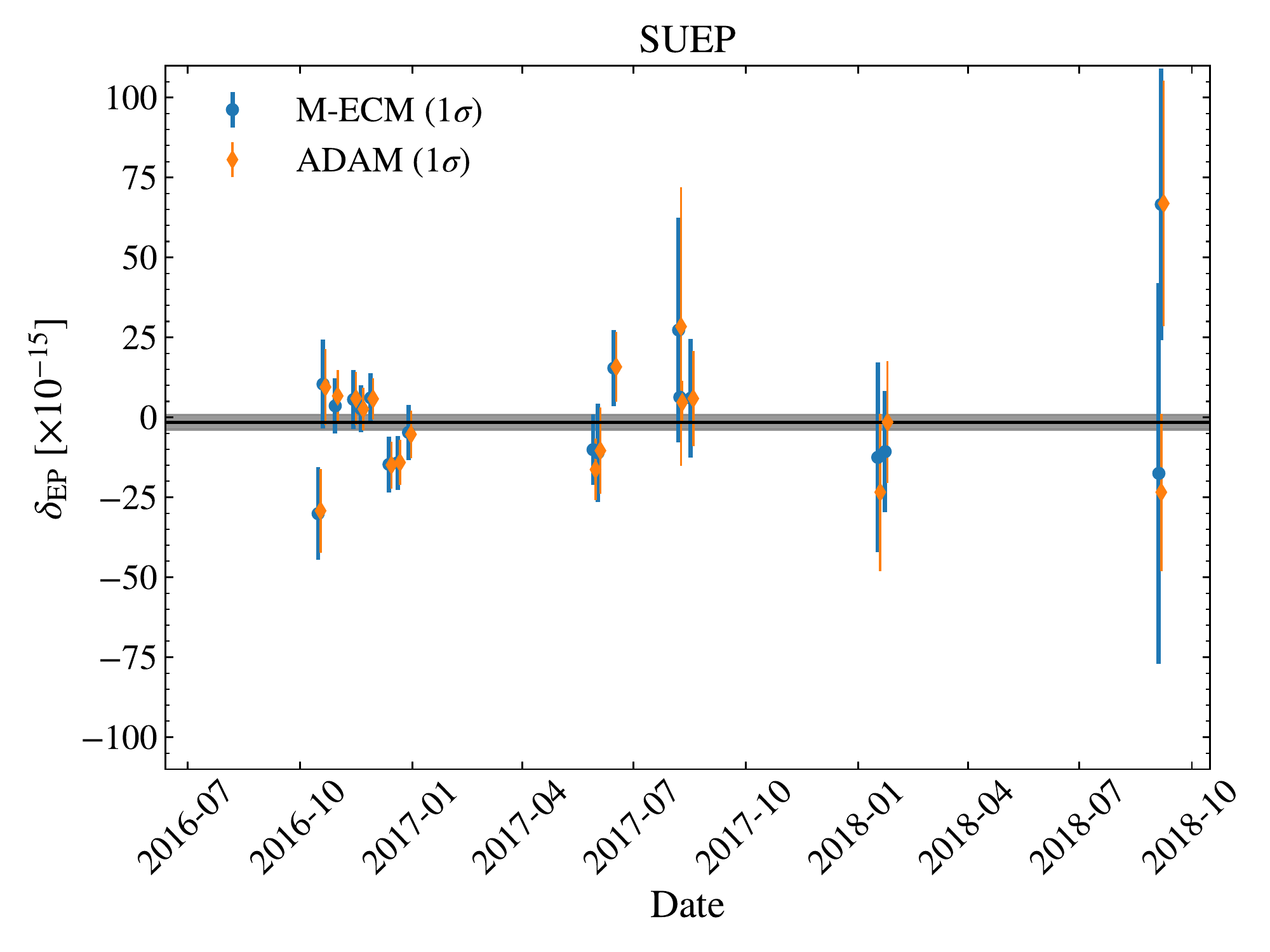}
\includegraphics[width=0.45\textwidth]{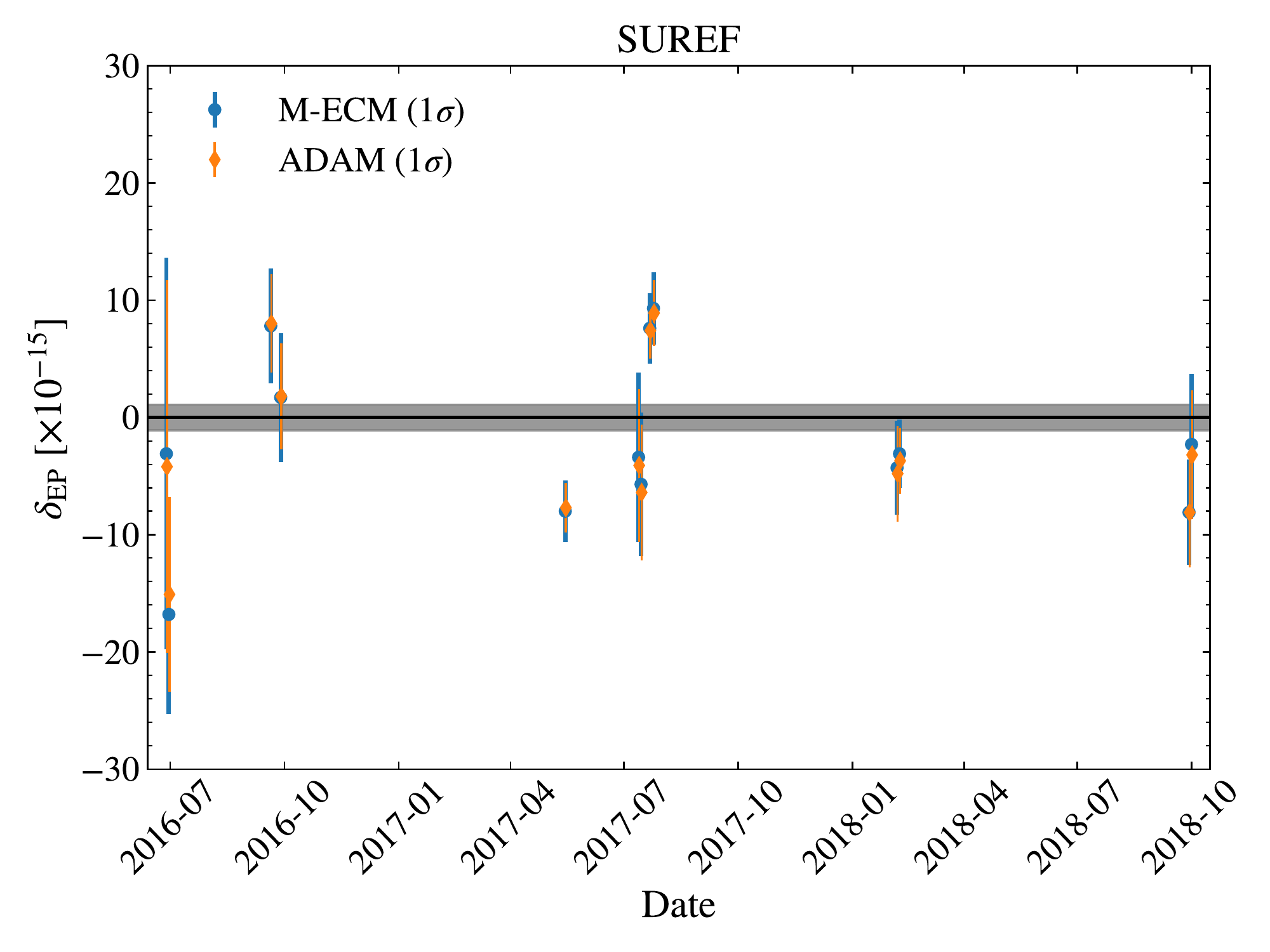}
\caption{E\"otv\"os parameter estimates for each segment and overall estimate and its 68\% confidence region (black line and grey area). Blue circles show M-ECM's estimates and orange ones {\sc Adam}'s. Upper panel: SUEP. Lower panel: SUREF. Panels span the same time.}
\label{fig_delta}
\end{figure}

Fig. \ref{fig_delta} shows the estimates of the E\"otv\"os parameter for each segment, obtained with M-ECM (blue circles) and {\sc Adam} (orange diamonds). The two methods are perfectly consistent. Error bars vary in accordance with the duration of segments and with the spin rate: the higher the spin rate, the lower the error bars, since the noise is minimal for the highest spin rate, see Ref. \cite{metriscqg9}. 
The black lines and grey areas show the combined constraints and their 68\% confidence region \cite{bergecqg7, metriscqg9}, $\delta({\rm Pt, Pt}) ~=~ (0.0\pm1.1)\times10^{-15}$ for SUREF and $\delta({\rm Ti, Pt}) ~=~ (-1.5\pm2.3)\times10^{-15}$ for SUEP. Those uncertainties contain statistical errors only. We discuss systematic errors below.

We found that the overall systematics upper bound is $2.3\times10^{-15}$ for SUREF and $1.5\times10^{-15}$ for SUEP, compared to specifications of $0.2\times10^{-15}$ \cite{hardycqg6}.
Except non-linearity (as discussed in Ref. \cite{chhuncqg5}) and temperature variations (see below), all contributors to the systematics error budget have effects lower than required. For instance, the contribution of the Earth gravity gradient could be cancelled by the precise estimation of the test masses offcenterings. Local gravity effects were mitigated by a careful design of the satellite and of the instrument \cite{hardycqg6}. Similarly, we showed in Ref. \cite{hardycqg6} that magnetic effects, due to the interaction of the test masses with the Earth magnetic field, were well below the requirements, as expected from the integration of a magnetic shield around the payload. The DFACS performance was much better than expectations, allowing for a residual linear accelerations at $f_{\rm EP}$ smaller than $2\times10^{-13}~{\rm ms}^{-2}$ (resp. $6\times10^{-14}~{\rm ms}^{-2}$) in (resp. out of) the orbital plane, and for residual angular accelerations smaller than $2.5\times10^{-11}~{\rm rad~s}^{-2}$; the related systematic errors are well below the specifications.

Temperature variations are the main source of systematic errors. They induce a differential acceleration through a thermal sensitivity of the SU and of the FEEU (Front End Electronics Unit).
Specific sessions were designed to characterize the thermal sensitivity through a periodic stimulus by on-board heaters; the temperature and differential accelerations were finely monitored and compared to provide better estimates of the thermal sensitivities at different frequencies \cite{hardycqg6, dhuicque21}. We found a linear frequency-dependence for SUEP’s thermal sensitivities, but none for SUREF’s.

On the other hand, the temperature data during EP sessions only allowed for a pessimistic upper bound since the temperature variations were smaller than the temperature probe’s noise at $f_{\rm EP}$.
In response to this limitation, additional sessions were devoted to confirm the thermal design of the satellite, in order to show that temperature variations are driven by the Earth's albedo coming through the FEEU radiator's baffle (Fig. 15 of Ref. \cite{hardycqg6}), inducing a modulation of the temperature at $f_{\rm EP}$ \cite{hardycqg6}.  A first session, based on heating the FEEU panel with local heaters, allowed us to show that the impact of the Earth’s albedo on the satellite walls is negligible.
In a second session, the satellite was tilted by 30$^{\rm o}$ about its spin axis in inertial mode during 465 orbits (32.3 days) in order to maximize the albedo light entering the FEEU radiator. 
We found that temperature variations are attenuated by a factor 500 between the FEEU and the SU.
Based on the data available (at frequencies lower than $10^{-3}~{}\rm Hz$), we took this factor 500 as the lowest limit to compute an upper bound of temperature fluctuations at frequencies higher than $10^{-3}~{\rm Hz}$, in particular at $f_{\rm EP}$, where temperature probes allow for a measurement of the FEEU temperature variations but not of the SU's, since it is below the probe's noise \cite{hardycqg6}.

\bigskip

Putting these results together, MICROSCOPE's new constraint on the validity of the WEP is
\begin{equation} \label{eq_eotvos}
\eta({\rm Ti, Pt}) = [-1.5 \pm 2.3~{\rm (stat)} \pm 1.5~{\rm (syst)}]\times 10^{-15},
\end{equation}
where the statistical error is given at $1\sigma$, and where we identified the measured, approximated E\"otv\"os ratio $\delta$ with the exact one $\eta$. This result is close to the $10^{-15}$ precision for which the mission was designed, and improves our previous constraints \cite{touboul17} by a factor 4.6. The reference instrument provided a null result, $\eta({\rm Pt, Pt}) ~=~ [0.0 \pm 1.1~{\rm (stat)} \pm 2.3~{\rm (syst)}]\times 10^{-15}$, showing no sign of unaccounted systematic errors in Eq. (\ref{eq_eotvos}). As expected, SUREF's statistical error is smaller than SUEP's because it is more sensitive.

Beside constraining the validity of the WEP to an unprecedented level, MICROSCOPE also allows for unprecedented constraints on topics as various as Lorentz invariance \cite{pihanlebar19}, long-range interactions \cite{berge18, fayet18, fayet19, pernotborras19, pernotborras20}, or dark matter searches \cite{graham16}.
It also paves the way to new, more ambitious experiments to test GR in space \cite{battelier21}. The analyses presented in this paper and in Refs. \cite{touboul17, touboul19} provide essential feedback for future upgrades on the payload and satellite sides that can lead to the next-generation MICROSCOPE mission. In particular, the gold wire allowing for the test masses charge management should be replaced by a contactless device, such as the one proven to work in space in LISA Pathfinder \cite{armano16, armano17, armano18}. Glitches should be reduced to a minimum, e.g. by tightening the requirement on the crackles of the satellite's coating, or their effect should be better understood, e.g. through a better understanding of the full transfer function of the satellite--instrument system, so that it can be efficiently corrected for. Furthermore, a better thermal stability and thermal characterization of the system will allow us to beat the thermal systematics. With these upgrades, it should be possible to reach a $10^{-17}$ precision on the E\"otv\"os ratio.

\bigskip
\noindent{\it Acknowledgements} The authors express their gratitude to all the different services involved in the mission partners and in particular CNES, the French space agency in charge of the satellite. This work is based on observations made with the T-SAGE instrument, installed on the CNES-ESA-ONERA-CNRS-OCA-DLR-ZARM MICROSCOPE mission. ONERA authors’ work is financially supported by CNES and ONERA fundings.
Authors from OCA, Observatoire de la C\^ote d'Azur, have been supported by OCA, CNRS, the French National Center for Scientific Research, and CNES. ZARM authors' work is supported by the DLR, German Space Agency,  with funds of the BMWi (FKZ 50 OY 1305) and by the Deutsche Forschungsgemeinschaft DFG (LA 905/12-1). The authors would like to thank the Physikalisch-Technische Bundesanstalt institute in Braunschweig, Germany, for their contribution to the development of the test-masses with funds of CNES and DLR.

\bibliography{micPRL21}
\end{document}